\documentclass[english,names,dvipsnames,DIV=14]{scrartcl}
\usepackage[T1]{fontenc}
\usepackage{mathpazo}
\usepackage{FiraSans}
\usepackage[ttdefault=true]{AnonymousPro} 
\usepackage{microtype}
\usepackage{amsmath,amssymb}
\usepackage{listings}
\usepackage{xcolor}
\usepackage{babel}
\date{}
\definecolor{RBackground}{HTML}{F8F9FA} 
\definecolor{RComment}{HTML}{6C757D}    
\definecolor{RKeyword}{HTML}{0056B3}    
\definecolor{RString}{HTML}{198754}     
\definecolor{RFrame}{HTML}{DEE2E6}      

\usepackage{hyperref}
\hypersetup{
    colorlinks=true,
    linkcolor=blue!20!black,
    filecolor=magenta!20!black,
    urlcolor=cyan!20!black,
    citecolor=green!20!black,
}
\newcommand{\email}[1]{\href{mailto:#1}{#1}}

\addtokomafont{captionlabel}{\bfseries}
\addtokomafont{caption}{\sffamily}
\setcapindent{0em}

\lstnewenvironment{CodeInput}{}{}
\lstnewenvironment{CodeOutput}{\lstset{language={}}}{}
\newenvironment{CodeChunk}{}{}

\newcommand{\pkg}[1]{\textbf{#1}}
\newcommand{\proglang}[1]{\texttt{#1}}
\usepackage{orcidlink,thumbpdf}

\usepackage{framed}

\author{Andrei Kramer\footnotemark[1]~\orcidlink{0000-0002-3828-6978} KTH Royal
  Institute of Technology, \and Federica Milinanni\thanks{These
    authors contributed equally as first
    authors.}~\orcidlink{0000-0003-3635-8760} ICERM,
  Brown University, \and Jeanette Hellgren
  Kotaleski~\orcidlink{0000-0002-0550-0739} KTH Royal Institute of
  Technology, \and Pierre Nyquist~\orcidlink{0000-0001-8702-2293}
  Chalmers University of Technology, \and Alexandra
  Jauhiainen\footnotemark[2]~\orcidlink{0000-0001-9714-7962} AstraZeneca \and Olivia
  Eriksson \thanks{These authors contributed equally as last authors.}
  ~\orcidlink{0000-0003-0740-4318} KTH Royal Institute of Technology }

\title{\pkg{uqsa}: an \proglang{R}-package for Bayesian uncertainty
  quantification and global sensitivity analysis for biochemical
  reaction network models}

\begin{document}
\maketitle
\begin{abstract}
   Biochemical reaction models describe dynamical subcellular
    processes. This model type is often subject to large, often
    unrecognized \emph{uncertainty}. To be able to account for this
    during the modeling process, we have developed the
    \proglang{R}-package \pkg{uqsa} for Bayesian uncertainty
    quantification and sensitivity analysis in an integrated
    fashion. The package is designed for fast simulations of dynamical
    models, and efficient sampling of high-dimensional parameter
    distributions, using Markov chain Monte Carlo (MCMC) sampling
    techniques, and global sensitivity analysis using two different
    methods, as well as local sensitivity estimation for efficient MCMC
    updates. We use Vine-copulas (via \pkg{VineCopula}) to model
    non-standard joint distributions. We provide builtin simulation
    tools for stochastic and deterministic models, in either
    likelihood-free or likelihood-based settings. \proglang{R} offers
    many more statistics tools for sample post processing, as well as
    packages that allow parallel execution that we interface with
    (\pkg{parallel}, \pkg{pbdMPI}).
\end{abstract}

\section{Introduction}\label{sec:intro}
Cellular processes, like cell division or synapse transmission, are
regulated by subcellular signaling pathways consisting of
\emph{networks of biochemical reactions}. It is important to learn the
dynamics of such networks in order to, e.g., predict the effects of
different drugs or disease mechanisms. This inference task is however
difficult, not only from a methodological but also from a practical,
technical perspective. The methodological part concerns the inference
of large, non-linear, mechanistic models with limited data. In
biological settings data acquisition is difficult, there are of course
limits to measurement techniques in experiments. This naturally
results in parameter estimates and predictions with a large
uncertainty. In many cases, this practical and/or structural
uncertainty has been ignored, partly because it is hard to quantify in
a statistically sound way. The technical difficulty relates to
challenges of handling large complicated network models and
transferring them to a mathematical formalism in a flexible,
interoperable and reusable way. We have extensively investigated both
of these challenges in previous work
\cite{eriksson2019uncertainty,santos2022modular,eriksson2022combining}
and present the \proglang{R}-package \pkg{uqsa} designed to address
them. The essential components of \textsc{uqsa}: \textbf{U}ncertainty
\textbf{Q}uantification and \textbf{S}ensitivity \textbf{A}nalysis are
combined with fast simulators, efficient and statistically sound
parameter estimation tools, embedded in a
\textsc{fair}\footnote{\textsc{fair}: Findable, Accessible,
  Interoperable, Reusable (guiding principles and requirements for the
  sharing of data and software)}\cite{wilkinson2016fair,
  eriksson2022combining} and user-friendly environment.

In Systems Biology, ill-posed inverse problems of this
type\footnote{parameter inference for mechanistic models given sparse
  data} have been extensively discussed from a range of perspectives
over the past decades \cite{cedersund2009systems,
  vanlier2013parameter}. Such models have been reported to exhibit
``sloppy'' parameter spaces \cite{gutenkunst2007universally} and to
be frequently non-identifiable \cite{raue2009structural,
  wieland2021structural}. Methods of model reduction have been
suggested \cite{eriksson2015modeling}.  However, as the models are
mechanistic, and the parameters therefore have a biological meaning,
reducing them to a fully identifiable form is not always
straightforward in a biologically meaningful way. We and others have
therefore pursued an alternative approach, namely to characterize and
quantify model uncertainty in relation to the available data so that
predictions can be accompanied by uncertainty estimates. Within
\pkg{uqsa} we also provide tools to analyze the sensitivity of
different model outputs to perturbations in model parameters. This
analysis can help guide the design of further experiments. In
addition, by representing model uncertainty explicitly, it becomes
possible to incorporate new data sequentially as they are generated,
thereby reducing uncertainty step by step, or alternatively revealing
deficiencies in the model structure.

In order to account for parameter uncertainty, we use a Bayesian
formalism. Rather than identifying \emph{one} unique optimal parameter
point during the parameter estimation process, this approach yields a
multivariate joint distribution of plausible parameter candidates
\cite{gelman1995bayesian, calvetti2023bayesian}. This uncertainty is
then propagated to the model predictions. This constitutes the UQ
(uncertainty quantification) part of \pkg{uqsa}. The other part is
global sensitivity analysis (SA) \cite{saltelli2008global}. Using
variance decomposition based methods, we can then attribute
uncertainty in the predictions (or other outputs) to uncertainty in
the different parameters~\cite{eriksson2019uncertainty}.  Sensitivity
analysis is also useful without data: we can analyses the relationship
between parameters and model output within the prior distribution,
which can be difficult to understand in complex models.

\begin{figure}
  \centering
  \includegraphics[width=\textwidth]{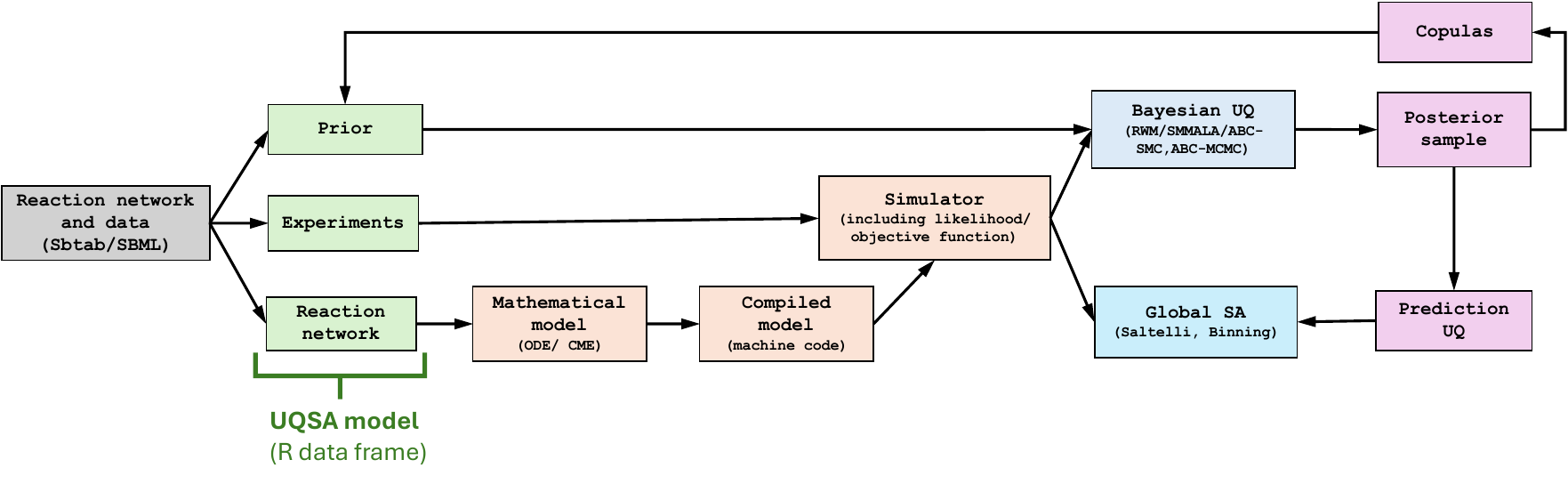}
  \caption{\label{fig:workflow}Schematic overview of the \pkg{uqsa}
    workflow. The \pkg{uqsa} package provides an integrated workflow
    for constructing, calibrating, and analyzing biochemical reaction
    network models. Models and experimental data are supplied through
    a systems biology format (gray) and transferred into \proglang{R}
    data frames (green). Models are next automatically translated into
    deterministic or stochastic mathematical representations (in
    \proglang{C}) that can be simulated and compared with data
    (orange). With the simulator, these models can (together with the
    prior) subsequently be subjected to Bayesian uncertainty
    quantification (dark blue) during calibration. The result is a
    posterior sample, which can be propagated to model predictions and
    described by a copula (all pink). Finally a global sensitivity
    analysis can be performed either on the posterior sample directly
    or from a prior sample without data restrictions.}
\end{figure}

Both \textsc{uq} and \textsc{sa} require a large number of
model simulations, and handling multidimensional parameter spaces in
an efficient and statistically appropriate way. \pkg{uqsa} combines
efficient simulators in \proglang{C} with preexisting advanced
statistical tools in \proglang{R} to model and calibrate dynamical
reaction network models. \pkg{uqsa} is sufficiently flexible to
accommodate both stochastic and deterministic
models. Figure~\ref{fig:workflow} shows a sketch of the workflow for
\pkg{uqsa}.

Other implementations of Bayesian uncertainty quantification for
similar types of models exist (e.g.,
\cite{klinger2018pyabc,schaelte2022pyabc,pyPESTO,GNUMCSim}), as do
implementations for global sensitivity analysis (e.g.,
\cite{tennoe2018uncertainpy}). Somewhat related projects also exist
for different research communities~\cite{STAN,NIMBLE}. However, to
our knowledge, there is no other tool for biochemical reaction
networks with such an extended and integrated repertoire. \pkg{uqsa}
performs both UQ and SA; it also offers the additional features of
automatic creation of deterministic or stochastic mathematical models
from reaction networks, and it allows the modeling of complex prior
distributions through Vine-copulas, and other \proglang{R} packages
related to distribution modeling and inference.

\pkg{uqsa} is available on GitHub under
\href{https://github.com/icpm-kth/UQSA}{\texttt{icpm-kth/UQSA}}.

\section{Theoretical background}

The main goal of the package \pkg{uqsa} is to estimate model
parameters and quantify their uncertainty in mathematical models of
biochemical reaction networks. To this end, we adopt two modeling
approaches for reaction networks: (i) describing the time evolution of
compound concentrations using ordinary differential equations (ODEs),
and (ii) describing the time evolution of the actual number of
molecules in the system using chemical master equations (CMEs).

\subsection{Mathematical models}

In this section, we introduce the ODE and CME modeling strategies.

\textbf{(I) ODE modeling of reaction networks.} Consider a biochemical
system with $n$ types of compounds $C_1,\dots,C_n$. The compounds
interact in the system via reactions, each of which is characterized
by a \textit{reaction rate coefficient}. These coefficients are
constants that describe how fast the corresponding reactions occur
based on the concentration of the reactants. Assuming that the
\textit{law of mass action} \cite{klipp2016systems, voit2015150}
holds, we can use these quantities to describe the time evolution of
the concentrations $[C_i]$ of each compound. For example, the
reversible reaction
$C_1+C_2\underset{k_b}{\stackrel{k_f}{\rightleftarrows }}C_3$ counts
as two reactions, where the forward reaction
$C_1+C_2{\stackrel{k_f}{\rightarrow }}C_3$ transforms the compounds
$C_1$ and $C_2$ into $C_3$ at a rate $k_f\cdot[C_1]\cdot[C_2]$, and
the backward reaction $C_1+C_2\underset{k_b}{{\leftarrow }}C_3$
transforms $C_3$ into $C_1$ and $C_2$ at a rate $k_b\cdot[C_3]$. If
the biochemical system consists only of these two reactions, we can
write the following ODE to describe the time evolution of the
concentrations $[C_1],[C_2],$ and $[C_3]$:
\begin{equation*}
    \begin{cases}
        \frac{d[C_1]}{dt}=-k_f[C_1][C_2]+k_b[C_3]\\
        \frac{d[C_2]}{dt}=-k_f[C_1][C_2]+k_b[C_3]\\
        \frac{d[C_3]}{dt}=+k_f[C_1][C_2]-k_b[C_3],
    \end{cases}
\end{equation*}
with given initial conditions $[C_i](t_0) = [C_i]_0$, for
$i=1,2,3$. This ODE system can be generalized to larger networks of
reactions with $n$ compounds $C_i$ as
\begin{equation*}
    \begin{cases}
        \dot{\mathbf{x}}(t)=\mathbf{f}(\mathbf{x}(t),\boldsymbol{\theta})\\
        \mathbf{x}(t_0)=\mathbf{x}_0,
    \end{cases}
\end{equation*}
where $\mathbf{x}(t)=(x_1(t),\dots,x_n(t))^T$ denotes the vector of
all compounds' concentrations $x_i(t)\in\mathbb{R}_{\ge0}$ at time
$t$, for example, the concentration of the different proteins and
protein complexes in the system, $\mathbf{x}_0\in\mathbb{R}_{\ge0}$ is
the vector of initial conditions for each compound, and
$\boldsymbol{\theta}$ is the vector of all reaction rate coefficients
(one for each reaction). The ODE vector field $\mathbf{f}$ may also
depend on controlled inputs $\mathbf{u}(t)$, e.g. calcium
concentration. In such case, we would write
$\dot{\mathbf{x}}(t)=\mathbf{f}(\mathbf{x}(t),\mathbf{u}(t),\boldsymbol{\theta})$.

The reaction rate coefficients in the vector $\boldsymbol{\theta}$ are
the model parameters that we aim to estimate. For this purpose, we use
data, which are \textit{noisy measurements} of a function
$\mathbf{g}(\mathbf{x}(t))$ of the state $\mathbf{x}(t)$ of the system
(i.e., concentrations) at given time points $t=t_1,\dots,t_K$. For
example, if only the first compound is measured, the function
$\mathbf{g}$ is defined as $\mathbf{g}(\mathbf{x})=x_1$ (recall that
$\mathbf{x}=(x_1,\dots,x_n)^T$). We also call the quantities
$\mathbf{g}(\mathbf{x}(t))$ the \textit{output} of the system at time
$t$.

We assume that the measurement $\mathbf{y}_k$ of the output
$\mathbf{g}(\mathbf{x}(t_k))$ at time $t_k$, $k=1,\dots,K$, is
characterized by additive Gaussian noise
$\mathcal{N}(\boldsymbol{0}, \Sigma_k)$, i.e.,
$\mathbf{y}_k\sim\mathcal{N}(\mathbf{g}(\mathbf{x}(t_k)),
\Sigma_k)$. Therefore, although the ODE model described above is
deterministic, the model that describes how data are generated is
stochastic. For this reason, we can define a likelihood function,
which is used in the Bayesian approach to uncertainty quantification
(see the next section for details).

\textbf{CME modeling of reaction networks.}  A limitation of the ODE
modeling approach is that the number of molecules of each compound
needs to be sufficiently large to be able to permit the definition of
compound \textit{concentrations}. This is not the case when we study
reaction mechanisms that take place in very small volumes. In such
situations, we instead model the reaction network using the chemical
master equation.

Within this approach, the state of the system is described by the
random vector $\mathbf{N}(t)=(N_1(t),\dots,N_n(t))^T$, whose
components $N_i(t)$ take values in $\mathbb{N}_{\ge0}$, and indicate
the \textit{random} number of molecules of compound $i$ in the system
at time $t$. The state $\mathbf{N}(t)$ changes in time due to the
random occurrence of reactions, which take place with a probability
related to the amount of reactants and to quantities called
\textit{propensities} depending on the state of the system. While in
the ODE modeling approach the state of the system $\mathbf{x}(t)$ is
deterministic and the randomness only lies in the measurement noise,
the state of the system $\mathbf{N}(t)$ is here intrinsically random.

Let $m$ be the number of possible reactions among the $n$ compounds
$C_1,\dots,C_n$. Assume that reaction $j$, $j=1,\dots,m$ changes the
state of the system from $\mathbf{n}$ to
$\mathbf{n}+\boldsymbol{\nu}_j$, where
$\boldsymbol{\nu}_j=(\nu_{j1},\dots,\nu_{jn})^T$ is a vector in
$\mathbb{R}^n$. For example, if $C_1+C_2\rightarrow C_3$ is the $j$-th
reaction, the components of $\boldsymbol{\nu}_j$ are $\nu_{j1}=-1$,
$\nu_{j2}=-1$, and $\nu_{j3}=+1$, and all other components are
$\nu_{j4}=\dots=\nu_{jn}=0$. If this reaction takes place at time $t$,
the state instantaneously changes from\footnote{$t^-$ indicates the
  time instant right before time $t$, thus,
  $\mathbf{N}(t^-)=\lim_{\tau\to t^-}\mathbf{N}(\tau)$.}
$\mathbf{N}(t^-)=\mathbf{n}$ to
$\mathbf{N}(t)=\mathbf{n}+\boldsymbol{\nu}_j$, i.e., the values of
$N_1$ and $N_2$ decrease by $1$, the value of $N_3$ increases by $1$,
and all other values $N_4,\dots,N_n$ remain unchanged.

Each reaction $j$ is characterized by a propensity $a_j$ depending on
the state of the system. The propensities $a_j$, $j=1,\dots,m$ are the
quantities that we want to estimate (and whose uncertainty we want to
quantify) in this modeling approach. These are defined so that the
quantity $a_j(\mathbf{n})\,dt$ corresponds to the probability that the
$j$-th reaction occurs in the infinitesimal interval $[t,t+dt)$ given
that the state of the system at time $t$ is
$\mathbf{N}(t)=\mathbf{n}$. Consequently, the probability evolution of
the state of the system $\mathbf{N}(t)$ is described by the
\textit{Chemical Master Equation}, or \textit{Kolmogorov forward
  equation},
\begin{equation*}
    \frac{\partial {P}(\mathbf{n},t|\mathbf{n}_0,t_0)}{\partial t} = \sum_{j=1}^m\left[a_j(\mathbf{n}-\boldsymbol{\nu}_j)\cdot P(\mathbf{n}-\boldsymbol{\nu}_j,t|\mathbf{n}_0,t_0) - a_j(\mathbf{n})\cdot P(\mathbf{n},t|\mathbf{n}_0,t_0)\right],
\end{equation*}
where $P(\mathbf{n},t|\mathbf{n}_0,t_0)$ is the probability that the
state of the system at time $t$ is $\mathbf{N}(t)=\mathbf{n}$, given
that the state at time $t_0$ is
$\mathbf{N}(t_0)=\mathbf{n}_0$. Solving the CME is usually unfeasible,
and therefore we can also not be able to define a likelihood function
for the Bayesian uncertainty quantification procedure. However, using
the \textit{Stochastic Simulation Algorithm} (SSA), also called the
\textit{Gillespie algorithm} \cite{gillespie2007stochastic}, we can
generate exact samples from the probability model for the time
evolution of $\mathbf{N}(t)$. Uncertainty quantification of the model
parameters, that is, the reactions' propensities $a_j$ in this case,
can next be performed with Approximate Bayesian Computation
techniques, which do not require formulation of a likelihood.  This
likelihood-free approach is described in a later section.

\textbf{Conservation laws} For a biochemical system where no
biochemical entities are added or removed, the concentration or number
of the entities must remain constant, even though they can be part of
different complexes. This law of mass conservation can be used to
simplify the system and remove linearly dependent differential
equations and replace them by appropriate algebraic equations
\cite{klipp2016systems}. Biochemical models also have to be consistent
with thermodynamical laws. This enforces dependencies between model
parameters necessary to consider for feasible models
\cite{klipp2016systems}.

\subsection{UQ - Bayesian parameter estimation}

Bayesian uncertainty quantification is a probabilistic framework for
estimating parameters while acknowledging their uncertainty
\cite{gelman1995bayesian}. In this framework, the (multivariate) model
parameter $\boldsymbol{\theta}$ is treated as a random variable
$\Theta$\footnote{We use capital letters to denote random variables,
  and lower case letters to indicate specific values.}. Bayesian UQ is
performed by defining the \textit{posterior} probability distribution
of the parameters conditioned on the observed data while taking into
account parameter prior information. By Bayes' theorem, the posterior
probability density can be expressed as the normalized product of the
\textit{prior} probability density of the parameters and the
\textit{likelihood} function. To characterize the posterior
distribution, we generate samples from it via Markov chain Monte Carlo
(MCMC) algorithms, considering both \textit{likelihood-based} and
\textit{likelihood-free} methods. In this section, we explain these
concepts in more detail.

\textbf{Prior.} The prior describes our assumptions on possible
parameter values before the parameter estimation process. The prior
density $\pi_{\text{pr}}(\boldsymbol{\theta})$ is thus the probability
density of the model parameter $\Theta$ before observing the
data\footnote{Throughout the paper, we will use the uppercase notation
  $\Theta$ to denote the random vector of model parameters, and the
  lower case $\boldsymbol{\theta}$ to indicate a specific value of the
  parameter vector, such as the argument in the prior density
  $\pi_{\text{pr}}(\boldsymbol{\theta})$, or samples
  $\boldsymbol{\theta}_i$ in the MCMC sampling (introduced
  later).}. We define $\pi_{\text{pr}}(\boldsymbol{\theta})$ based on
prior information obtained, e.g., from the literature. For example, if
we already have a rough estimate $\boldsymbol{\theta}_{\text{pr}}$ for
$\Theta$, we could define $\pi_{\text{pr}}(\boldsymbol{\theta})$ as
the probability density function of a normal distribution centered at
$\boldsymbol{\theta}_{\text{pr}}$, with sufficiently large
variance. In \pkg{uqsa}, the prior
$\pi_{\text{pr}}(\boldsymbol{\theta})$ is defined based on information
provided in the model tables (TSV).

\textbf{Likelihood.} The likelihood function
$\pi_{\text{lik}}(\mathbf{y} \mid \boldsymbol{\theta})$ measures how
well the model with parameter $\boldsymbol{\theta}$ fits the data
$\mathbf{y}$. Let $\mathcal{D}_{\theta}(\cdot)$ be the probability
distribution of the data $\mathbf{y}$ depending on the parameter
$\boldsymbol{\theta}$. That is, the data $\mathbf{y}$ are seen as
realizations of the random variable
$\mathbf{Y}\sim\mathcal{D}_{\theta}(\cdot)$, for a fixed model
parameter $\boldsymbol{\theta}$. The likelihood
$\pi_{\text{lik}}(\mathbf{y} \mid \boldsymbol{\theta})$ corresponds to
the probability density of $\mathcal{D}_{\theta}(\cdot)$ for a fixed
$\boldsymbol{\theta}$. Note, however, that the likelihood is viewed as
a function of the parameter, that is,
$\boldsymbol{\theta} \mapsto \pi_{\text{lik}}(\mathbf{y} \mid
\boldsymbol{\theta})$. The observations $\mathbf{y}$ are assumed to be
known and are therefore treated as constants.

In \pkg{uqsa}, we can easily define the likelihood function when the
data are modeled using a deterministic model together with an assumed
additive Gaussian measurement noise. For example, we may model the
time evolution of protein concentrations in an experimental setting
using ordinary differential equations, and assume that the observed
data correspond to noisy measurements of a function of these
concentrations at given time points.

When the data are treated as realizations of a stochastic model, a
likelihood function is usually hard to determine. In such cases, we
recommend using likelihood-free approaches (see below).

\textbf{Posterior.} The posterior density
$\pi_{\text{pos}}(\boldsymbol{\theta} \mid \mathbf{y})$ is the
probability density of the model parameter $\Theta$ after observing
the data $\mathbf{y}$. By Bayes' theorem, it is given by
\begin{equation*}
  \pi_{\text{pos}}(\boldsymbol{\theta}|\mathbf{y})=\frac{ \pi_{\text{pr}}(\boldsymbol{\theta})\pi_{\text{lik}}(\mathbf{y}|\boldsymbol{\theta})}{\int\pi_{\text{pr}}(\boldsymbol{\vartheta})\pi_{\text{lik}}(\mathbf{y}|\boldsymbol{\vartheta})d\boldsymbol{\vartheta}}.
\end{equation*}
The integral in the denominator is difficult to compute, especially in
high dimensions. For this reason, \pkg{uqsa} implements MCMC
algorithms to generate samples from the posterior
$\pi_{\text{pos}}(\boldsymbol{\theta} \mid \mathbf{y})$ as such
methods only need the prior $\pi_{\text{pr}}(\boldsymbol{\theta})$ and
the likelihood
$\pi_{\text{lik}}(\mathbf{y} \mid \boldsymbol{\theta})$. In addition,
\pkg{uqsa} implements an Approximate Bayesian Computation MCMC method,
which is likelihood-free and can therefore be used in cases where the
likelihood is not available.

\textbf{Likelihood-based sampling.}  In \pkg{uqsa}, MCMC algorithms
are used to generate samples that are approximately distributed
according to the posterior distribution
$\pi_{\text{pos}}(\boldsymbol{\theta} \mid \mathbf{y})$. In
particular, we consider algorithms of \textit{Metropolis-Hastings}
type \cite{robert2004monte}. These algorithms generate a Markov chain
of samples $\boldsymbol{\theta}_1,\boldsymbol{\theta}_2,\dots$ as
follows: at each iteration,
\begin{enumerate}
\item given the current sample $\boldsymbol{\theta}_i$, a proposed
  value $\boldsymbol{\theta}^\star$ for the $(i+1)$-th term of the
  chain is drawn from a \textit{proposal probability distribution}
  $q(\cdot|\boldsymbol{\theta}_i)$ with density
  $\boldsymbol{\theta}\mapsto
  q(\boldsymbol{\theta}|\boldsymbol{\theta}_i)$ that depends on
  $\boldsymbol{\theta}_i$,
\item the proposed sample $\boldsymbol{\theta}^\star$ is then accepted
  (i.e., we set $\boldsymbol{\theta}_{i+1}=\boldsymbol{\theta}^\star$)
  with probability \eqref{eq:acceptance}, otherwise, it is rejected
  (i.e., we discard $\boldsymbol{\theta}^\star$, and set
  $\boldsymbol{\theta}_{i+1}=\boldsymbol{\theta}_i$).
\end{enumerate}

\begin{equation}
  \label{eq:acceptance}
  P_{\text{acc.}}=\min\left\{1,\frac{
      \pi_{\text{pos}}(\boldsymbol{\theta}^\star)q(\boldsymbol{\theta}_i|\boldsymbol{\theta}^\star)
    }{
      \pi_{\text{pos}}(\boldsymbol{\theta}_i)q(\boldsymbol{\theta}^\star|\boldsymbol{\theta}_i)
    }\right\}
\end{equation}

Different choices of the proposal distribution
$q(\cdot \mid \boldsymbol{\theta}_i)$ lead to different
Metropolis-Hastings algorithms. The main algorithms implemented in
\pkg{uqsa} are \textit{Random Walk Metropolis} (RWM) and the
\textit{Simplified Manifold Metropolis-adjusted Langevin Algorithm}
(SMMALA), which are characterized by the following two proposal
distributions: $q(\cdot|\boldsymbol{\theta}_i)$ is
$\mathcal{N}(\boldsymbol{\theta}_i,\Sigma)$ for RWM, where $\Sigma$ is
a given covariance matrix, and
$\mathcal{N}(\boldsymbol{\theta}_i +
\frac{h}{2}G(\boldsymbol{\theta}_i)^{-1}\nabla_{\boldsymbol{\theta}}\pi_{\text{pos}}(\boldsymbol{\theta}_i|y),h
\,G(\boldsymbol{\theta}_i)^{-1})$ for SMMALA, where $h$ is the
method's step-size and $G(\boldsymbol{\theta})$ is the expected Fisher
information matrix \cite{Girolami2011}.

\textbf{(II) CME models and Likelihood-free sampling.}  When the exact
mathematical likelihood is unavailable due to complex model
stochasticity, we can rely on \emph{likelihood-free} sampling
algorithms, that allow us to generate samples approximately
distributed as
$\pi_{\text{pos}}(\boldsymbol{\theta}|\mathbf{y})$. However, it is
important to emphasize here, that this offers no computational benefit
compared to methods with a known likelihood function. The ABC method
requires a defined \emph{distance} function, which plays the same role
as a likelihood function\footnote{for deterministic models} with
roughly the same costs. In fact, a stochastically embedded
deterministic ODE model\footnote{stochastically wrapped: a noise
  process is added to the model's otherwise deterministic output (or
  applied multiplicatively)} with ABC and a deterministic ODE model
with Metropolis MCMC both have to be simulated (once each) and have
the same computational burden\footnote{the stochastic embedding is
  negligible in computational costs, compared to ODE solvers}.

The likelihood-free algorithm that we provide in \pkg{uqsa} is the
\textit{Approximate Bayesian Computation} MCMC (ABC-MCMC) method
\cite{Marjoram2003}, and ABC-SMC. ABC-MCMC requires the availability
of a simulator that can generate data from the stochastic model for
the data for a given parameter value $\boldsymbol{\theta}$, i.e., we
can generate
$\mathbf{y}^{\text{sim}}\sim\mathcal{D}_{\boldsymbol{\theta}}(\cdot)$
\footnote{Recall that $\mathcal{D}_{\boldsymbol{\theta}}(\cdot)$ is
  the probability model for the data for fixed parameter
  $\boldsymbol{\theta}$, and
  $\pi_{\text{lik}}(\mathbf{y}\mid \boldsymbol{\theta})$ is the
  corresponding probability density}.  Note that this approach does
not necessitate the availability of the likelihood function
$\mathbf{y}\mapsto \pi_{\text{lik}}(\mathbf{y}\mid
\boldsymbol{\theta})$. The algorithm is implemented as follows: at
each iteration,
\begin{enumerate}
\item given the sample $\boldsymbol{\theta}_i$, a value
  $\boldsymbol{\theta}^\star$ for the $(i+1)$-th term of the chain is
  drawn from a \textit{proposal probability distribution}
  $q(\cdot|\boldsymbol{\theta}_i)$ with density
  $\boldsymbol{\theta}\mapsto
  q(\boldsymbol{\theta}|\boldsymbol{\theta}_i)$ that depends on
  $\boldsymbol{\theta}_i$;
\item the proposed sample $\boldsymbol{\theta}^\star$ is plugged in
  into the stochastic model for the data to generate simulated data
  $\mathbf{y}^{\text{sim}}$ (that is,
  $\mathbf{y}^{\text{sim}}\sim\pi_{\text{lik}}(\cdot|\boldsymbol{\theta}^\star)$);
\item given a distance measure $\rho(\cdot,\cdot)$ on the data space
  and a threshold $\varepsilon>0$,
  \begin{itemize}
  \item if $\rho(\mathbf{y},\mathbf{y}^{\text{sim}})\ge\varepsilon$,
    we reject the proposal $\boldsymbol{\theta}^\star$, and set
    $\boldsymbol{\theta}_{i+1}=\boldsymbol{\theta}_i$
  \item otherwise, we proceed with a Metropolis-Hastings
    acceptance-rejection step, accepting the proposal
    $\boldsymbol{\theta}^\star$ with probability
    \eqref{eq:acceptance-2} (thus, set
    $\boldsymbol{\theta}_{i+1}=\boldsymbol{\theta}^\star$),
    otherwise, $\boldsymbol{\theta}_{i+1}=\boldsymbol{\theta}_i$.
  \end{itemize}
\end{enumerate}

\begin{equation}
  \label{eq:acceptance-2}
  P_{\text{acc.}}^{MH}= \min\{1,\frac{
    \pi_{\text{pr}}(\boldsymbol{\theta}^\star)q(\boldsymbol{\theta}_i|\boldsymbol{\theta}^\star)
  }{
    \pi_{\text{pr}}(\boldsymbol{\theta}_i)q(\boldsymbol{\theta}^\star|\boldsymbol{\theta}_i)
  }
  \}
\end{equation}
We recommend using a likelihood-based scheme (e.g., RWM or SMMALA)
when the likelihood function
$\pi_{\text{lik}}(\mathbf{y} \mid \boldsymbol{\theta})$ is available,
as these methods generate samples that more accurately approximate the
posterior $\pi_{\text{pos}}(\boldsymbol{\theta} \mid
\mathbf{y})$. ABC-MCMC requires very little information, e.g.: no
objective-function gradients. We consider \emph{simplicity} as one of
the advantages of ABC methods, and use fairly simple proposal
mechanisms, such as the RWM proposal, for which
$q(\cdot \mid \boldsymbol{\theta}_i)$ is
$\mathcal{N}(\boldsymbol{\theta}_i,\Sigma)$ for a given covariance
matrix $\Sigma$.

For smaller models, it is feasible to use ABC-SMC, a particle
filter. In ABC-SMC, we operate on a whole sample of points. We start
with a prior sample and each iteration of the particle filter moves
this sample closer to the posterior, until it converges. The output is
either a weighted sample (with weights indicating important
parameter-space locations), or a classic sample where a good fit is
indicated by a large number of sampled points.

\subsection{SA - variance decomposition based}

With \pkg{uqsa}, global sensitivity analysis (GSA) can be performed to
investigate how uncertainty in the model output depends on uncertainty
in the parameter estimates. The GSA implemented in the package is
based on variance decomposition. For general introductions to GSA,
see, e.g., \cite{saltelli2008global, zi2011sensitivity,
  tennoe2018uncertainpy}.

\textbf{Sobol sensitivity indices.} The sensitivity can be described
by the Sobol sensitivity indices \cite{sobol2001global}.  There are
several types of Sobol indices. The \emph{first order} sensitivity
index $S_i$ measures the direct effect that a parameter $\Theta_i$ has
on the uncertainty of the output of the model
\cite{saltelli2008global}, i.e., how much of the variance of the
model output $\text{Var}(Y)$ that can be explained by $\Theta_i$ alone
\begin{equation}
S_i=\frac{\text{Var}(\mathbb{E}(Y|\Theta_i))}{\text{Var}(Y)}.
\end{equation}
Here, $\mathbb{E}(Y|\Theta_i)$ denotes the expected value of the
output $Y$ when $\Theta_i$ is fixed, and for varying values of all
other parameters, and the variance is over all possible values of
fixed $\Theta_i$ (see, e.g., chapter 5 of
\cite{saltelli2004sensitivity} for a detailed explanation). Two
parameters are said to \emph{interact} if their combined effect on $Y$
cannot be expressed as the sum of their single effects on $Y$
\cite{saltelli2004sensitivity}. Also higher order effects can be
described with this approach. Usually, one calculates the the first
order effects as well as the \emph{total order} effects. The total
order sensitivity index of a specific parameter, $S_{Ti}$ includes the
first order effects and all possible combinations of interactions
between this parameter and all other parameters, and is calculated by

\begin{equation}
  S_{Ti}=1-\frac{\text{Var}(\mathbb{E}(Y|\Theta_{-i}))}{\text{Var}(Y)}
\end{equation}
where $\Theta_{-i}$ denotes all parameters except $\Theta_i$.

\textbf{Methods for Orthogonal or non-orthogonal distributions.}  The
classical case for GSA is that the parameter distributions are
orthogonal. In that case, more efficient methods can be used to
calculate both the first-order sensitivity indices, $S_i$, and the
total order sensitivity indices, $S_{Ti}$. For this setting,
\pkg{uqsa} implements the Saltelli method \cite{saltelli2002making},
following \cite{halnes2009modelling}. In some cases, it is of interest
to compute the sensitivity on a posterior sample, for example to
determine which single parameter contributes most to the uncertainty
in a prediction. Such a parameter would be an important target for
experiments aim at reducing the uncertainty in the predictions. In
most systems biology settings, however, the posterior sample is
non-orthogonal, as dependencies between parameters are reflected by
the model structure. In this case, the Saltelli method cannot be
used. Instead, we have implemented a method for calculating first
order sensitivity indexes, $S_i$, called the binning-method (for
details, see supplementary material of
\cite{eriksson2019uncertainty}), in which the posterior distribution
is subdivided into bins and $S_i$ is calculated. Beyond answering the
question of which single parameter contributes most to prediction
uncertainty, the results of a GSA study with a non-orthogonal
distribution are generally harder to interpret
\cite{saltelli2004sensitivity} because of potential dependencies
between parameters.

\subsection{Copulas}
\label{sec:copulas}

Copulas flexibly model multivariate distributions by separating
marginal distributions from the dependence structure. Let
$\boldsymbol{X} = (X_1, \ldots, X_d)$ have continuous marginals
$F_i(x_i) = P(X_i \le x_i)$. By Sklar’s theorem, there exists a unique
copula $C$ such that \begin{equation*} F(x_1, \ldots, x_d) =
  C\big(F_1(x_1), \ldots, F_d(x_d)\big), \end{equation*} so
$U_i = F_i(X_i)$ are $\mathrm{Uniform}(0,1)$ and $C$ captures
dependence. The joint density factorizes into marginal densities times
a copula density, which is advantageous with e.g. non-Gaussian
marginals or nonlinear dependencies.

Pair-copula constructions build high-dimensional dependencies from
bivariate copulas organized with graphical models called
vines. Regular vines (R-vines) allow flexible structures to organize
the copulas. Practically, one selects pair-copula families (e.g.,
Gaussian, Clayton, Gumbel, Frank, Joe, or rotated variants) and
estimates marginal distributions and copula parameters.

In \pkg{uqsa}, copulas can be used to model complex, non-orthogonal
posterior parameter distributions in sequential dataset fitting: an
R-vine copula model fitted to a posterior sample describes the viable
parameter space and can be used to form the prior for the next
dataset. This preserves learned dependencies and avoids unrealistic
independence assumptions.

\subsubsection{Parallel Tempering}
\label{sec:parallel-tempering}

The parallel tempering method deals
with an ensemble of several Markov chains with different
\emph{temperatures} $\beta_i\in [0,1]$ \cite{Earl2005,Miasojedow2013}:
\begin{equation}
  \label{eq:temperatures}
  p(\theta|y)_i = \frac{
    p(y|\theta)^{\beta_i} p(\theta)
  }{
    \int_{\mathfrak{T}} p(y|\theta)^{\beta_i}p(\theta) d\theta
  }\,.
\end{equation}
where $i$ enumerates the different Markov chains (and temperatures),
$\mathfrak{T}$ is the domain of $\theta$. Any temperature $\beta$
lower than $1$ effectively relaxes the problem: a value of $0$ results
in a sampling from the prior distribution, a value of $1$ results in a
conventional posterior sample (as described in previous
sections). Values in-between ($0<\beta_i<1$) sample from intermediate
distributions. The different chains can swap their positions under the
right circumstances (or equivalently, their temperatures). This mixing
between differently relaxed modes of operation dramatically speeds up
conversion to the target distribution\footnote{decreases auto-correlation}.

\section{Software}
\label{sec:software}

In this section we describe the features and design goals of
\pkg{uqsa}. Next we describe our approach to parallelization and HPC
applications.

One of the major design goals was to create simple intermediate files
that allow alternative workflows to be injected into ours. Even though
we envision this package to be used with systems biology models, it is
possible to use any ODE model, as long as it is written for the
solvers in the GNU Scientific Library (GSL) \cite{GSL,
  galassi1996gnu}.  It is also possible to use any R model, if the
user supplies their own custom likelihood, objective function, or
distance function.

\subsection{Features}
\label{sec:features}

The primary form of model in systems biology is at its \emph{core} a
list of reactions, with reaction fluxes\footnote{kinetic laws} with
reacting compounds and parameters that govern the reaction speed. We
record such a model as a list of TSV files, where each file tabulates
the entities of one type. It is the most natural way to store this
information, and TSV files are uniquely suited for this task. They
require no special parser, are easy to track with version control
systems, and require no quoting as tabs are never part of any field's
content.  It is also trivially easy to add tabular information such as
data, or a series of experimental setups to recreate these data-sets
using the given model.

In addition to the \emph{core} model, we also specify an
\emph{observation model}\footnote{also referred to as: \emph{output
    functions}, or \emph{model output}}, which connects internal model entities
(compounds) to the recorded measurements of the real system.

The plain tabular model specification format that we use is very
similar to SBtab (systems biology tables) format
\cite{lubitz2016sbtab}. We offer a companion package \pkg{SBtabVFGEN},
which processes SBtab files and generates ODE models in various
formats:
\href{https://github.com/icpm-kth/SBtabVFGEN}{\texttt{icpm-kth/SBtabVFGEN}}

A widely accepted standard format in the field is
SBML\footnote{Systems Biology Markup Language}, these models can be
translated into SBtab using the official tools on~\url{sbtab.net},
including web-hosted tools.

\textbf{Reading reactions, experiments and prior - from text files into \proglang{R} data frames}

The TSV files with constants, parameters (kinetic rates), intermediate
expressions, reactions, model outputs, list of experiments, state
transformations, event schedules, and data sets are read into
\proglang{R} as a list of named data frames. There is minimal
interpretation of the contents while reading the files.

The data-sets are imported as a list of simulation experiments, with
specific initial values, inputs, and measurement time values (and
optionally scheduled events) needed to reproduce the equivalent of the
data within the model. The data-sets are stored as matrices, with rows
corresponding to time-series of one named output function. The row
order is the same as in the output table (\texttt{Output.tsv}).

Data values with parenthetical error notation are interpreted
correctly, e.g.: \texttt{12.3(1)} is interpreted as $12.3$ with a
standard error of $0.1$. We use the \pkg{errors} package to store this
information.

\textbf{Simulating reaction networks as deterministic or stochastic
  mathematical models and comparing with data} A reaction network
model can be automatically translated into either a stochastic
chemical master equation (CME) or an ordinary differential equation
(ODE). A corresponding \emph{simulator} is available for each model
type. The simulators are written in \proglang{C} in order to be
computationally efficient.

When performing inference, a model is simulated many times. For this
reason we create the simulator as a closure (\proglang{R} function with
an internal data environment) constructed from the \emph{mathematical
  model} and the list of simulation \emph{experiments}.

In this context, an \emph{experiment} is a \emph{data set} and the
complete known set-up for the solver that is suitable to reproduce the
data using a simulation, in principle, within the limits of our
knowledge. Each ongoing modeling/inference task has unknown
components, in our case the unknowns are the model parameters
(e.g. kinetic rate coefficients). For this reason we don't include the
kinetic parameters in the list of experiments: the parameter vector is
the only free argument to the simulation closure; the model and list
of experiments are \emph{implicitly} stored in the simulator, after
its creation.

We use discrete \emph{scheduled events} to represent experimental
input protocols. This is particularly important in neuroscience
applications, where sudden, transient, time-dependent inputs such as
\emph{spike trains} are very common.

While our built-in simulators are written in \proglang{C} for
efficiency, it is also possible to use other R solvers, e.g.: from the
\pkg{deSolve}~\cite{deSolve} and
\pkg{GillespieSSA2}~\cite{Canoodt2021} packages, by writing a custom
likelihood or distance function.

\begin{figure*}
\centering
\includegraphics[width=\textwidth]{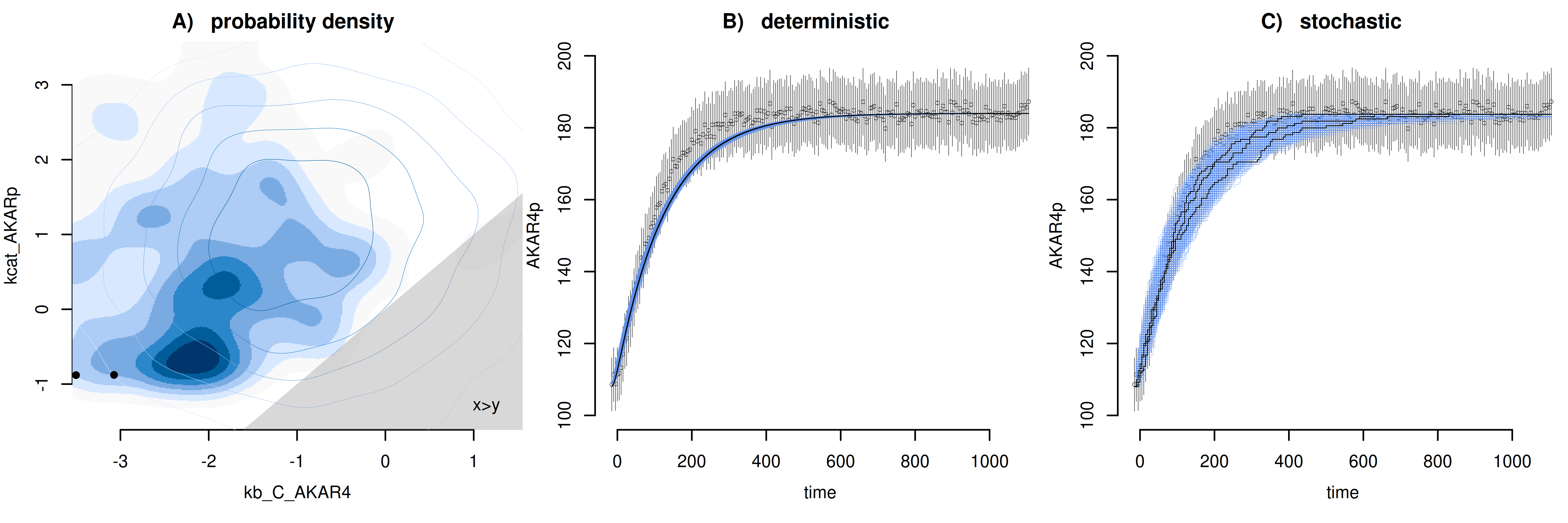}
\caption{\emph{Bayesian uncertainty quantification.} A) Prior and
  posterior distributions projected onto two parameter components. The
  prior is illustrated by contour lines and the posterior by a cloud
  of samples. The maximum likelihood estimate is shown as a point. B)
  A subset of the experimental data used for fitting and the
  corresponding model output (AKAP4p). The experimental data are
  represented by black circles, with vertical bars indicating
  measurement errors. Solid curves correspond to outputs from the
  fitted model: the black curve corresponds to the model with the MLE
  parameter values, while the
  \textcolor{CornflowerBlue}{cornflower-blue} curves correspond to the
  model with parameter values sampled from the posterior
  distribution. The spread of the blue curves indicates the
  uncertainty. C) The same model simulated stochastically (Gillespie
  algorithm). The transparent light blue lines correspond to samples
  from the posterior and represent the uncertainty in the
  prediction.\label{fig:UQ}}
\end{figure*}

\textbf{Uncertainty Quantification (UQ).}  UQ is performed during the
parameter estimates using a Bayesian approach through Markov chain
Monte Carlo (MCMC) methods \cite{gelman1995bayesian}. We sample from
the Bayesian posterior distribution of model parameters given
experimental data, and thereby obtain a representative sample of the
parameter region that provides a good fit to the data
(Figure~\ref{fig:UQ}). We have implemented a number of MCMC methods,
suited for different models and scenarios, including sampling from
deterministic models with additive measurement noise, where the
likelihood is easily computed, or from stochastic models for which the
likelihood cannot be directly calculated.  We have implemented two
different likelihood-based algorithms: (i)~the Random Walk Metropolis
(RWM) algorithm \cite{mengersen1996rates,Hastings1970, Gelman1997}, a
standard method that can be used for most models but can suffer from
high auto-correlation, and (ii) the Simplified Manifold
Metropolis-Adjusted Langevin Algorithm (SMMALA)
\cite{GirolamiMark2011RmLa, kramer2010computation}, a
geometry-informed algorithm with a fairly high level of
complexity. SMMALA is faster than RWM, but can be difficult to use in
some situations. We use both approaches within a parallel tempering
scheme \cite{Earl2005,swendsen1986replica}, which speeds up the
sampling process by running multiple MCMC chains at different
temperatures (levels of problem relaxation) and allowing the chains to
swap positions when beneficial. This increases the speed of parameter
space exploration. For cases where the likelihood is intractable or
complex, as for stochastic models, we provide an Approximate Bayesian
Computation (ABC) MCMC scheme \cite{Marjoram2003}, either plain
ABC-MCMC or ABC using sequential Monte Carlo, ABC-SMC
\cite{Sisson2007,toni2009approximate}. In SMC a population of random
samples (particles) are used to represent the entire probability
distribution of the system's true state.

By default we use an ABC distance function in the form of weighted
Euclidean distance, with weights given by (the reciprocals of) the
experimental measurement errors (when
available)~\cite{Prangle2017}. The ABC distance function can be
defined by the user. The default likelihood is calculated during the
simulation assuming Gaussian measurement errors.

\begin{figure*}
\centering
\includegraphics[width=\textwidth]{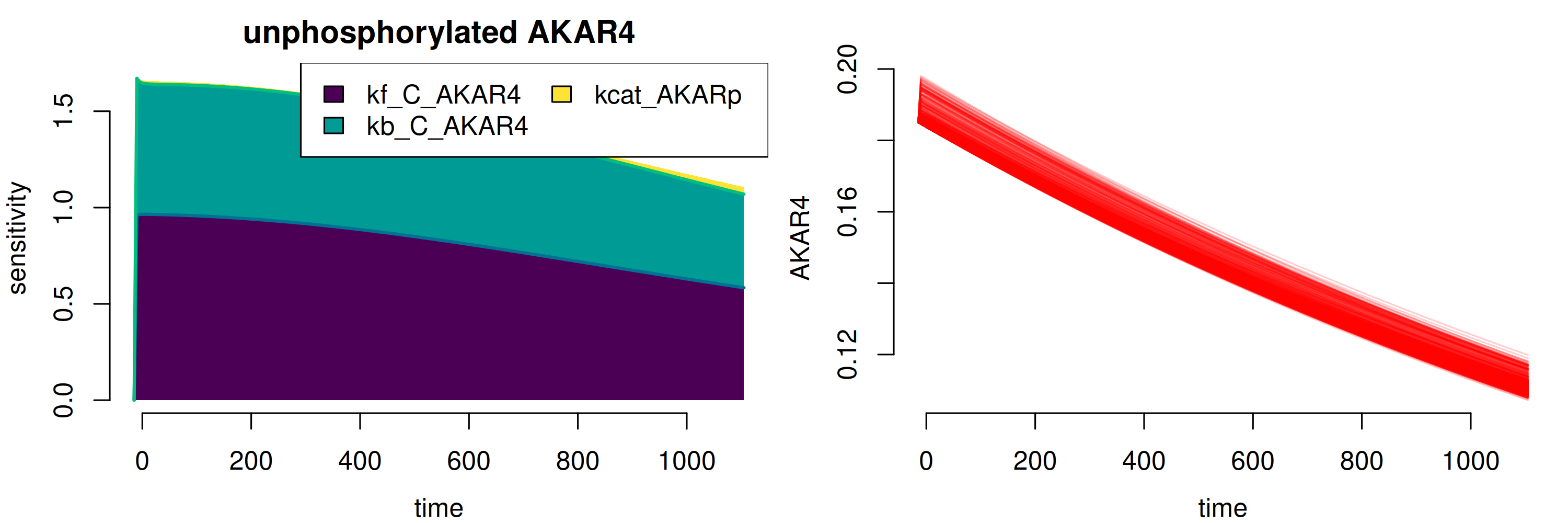}
\caption{ \emph{Prediction:} We created a simulation experiment with
  an input of $C:=0.01$, a setting for which no data has been
  collected yet. We show the sensitivity of unphosphorylated AKAR4, a
  state variable that was not observed in any of the experiments, and
  isn't necessarily constrained by the available data.  \emph{(left)}
  Global sensitivity analysis using the binning method, shown as
  cumulative shaded areas. \emph{(right)} forward simulation of
  unphosphorylated AKAR4, from sampled parameters.} \label{fig:SA}
\end{figure*}

\textbf{Merging datasets through Vine-Copulas.} A huge benefit of
working in R is a rich package-library for the modeling of probability
distributions. We use the VineCopula package to model complex joint
distributions with very high fidelity. When fitting a model, there are
often multiple datasets derived from different experimental set-ups,
with different read-outs (observables), to consider. For example, we
may consider mutations in which one reaction can be turned on or off,
or experimental data that consist of time series or different
steady-state outputs. In some cases the experiments are performed at
different times and are not processed on the same day.

There are different ways to combine these data-sets.  In \pkg{uqsa},
all experiments can be considered at once during the MCMC
sampling. But, a deliberate alternative is to use the experimental
datasets one after another. In this case, the posterior samples
obtained by fitting one dataset are used to formulate a prior
distribution for fitting the subsequent dataset
\cite{eriksson2019uncertainty}. This procedure requires modeling the
often high-dimensional posterior distribution of the sampled
parameters (probability density inference).

Our approach is to use multivariate probability distributions called
copulas, linked via graphical models referred to as vines, thereby
forming \emph{vine-copulas} \cite{Brechmann2013,Nagler2025}
(\pkg{VineCopula}). Copulas enable the modeling of multivariate
probability distributions by separating the marginal distributions
from the dependence structure of the stochastic variables. Vines can
be used to formulate copulas that are constructed pairwise in order to
describe dependencies across multiple variables
\cite{bedford2002vines}.

\begin{figure}
  \centering
  \includegraphics[width=\linewidth]{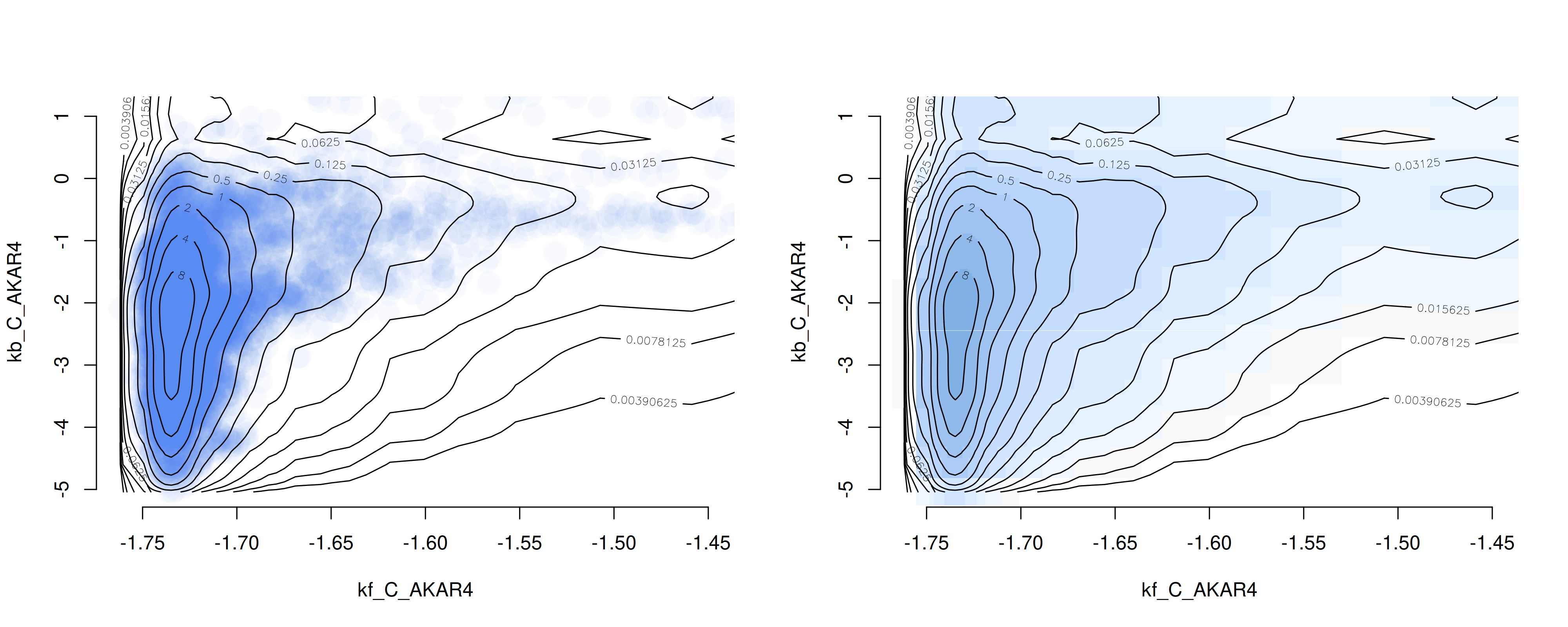}
  \caption[scatterplot and density plot]{A comparison between a 2D
    scatter plot and the density estimate via \pkg{VineCopula}. We
    used the first two columns (two parameters) of an MCMC sample
    ($N=8696$) for the AKAR4 model. The left panel shows a
    scatterplot, with sample members being represented by filled
    circles with very low opacity (where circles accumulate, their
    opacity adds up). The second plot shows a raster image where we
    evaluated the vine-copula estimate of the density on grid points.}
\end{figure}

\textbf{Global sensitivity analysis on orthogonal and non-orthogonal
  distributions.} Global sensitivity analysis can be used to
investigate how different input factors (kinetic parameters in our
case) affect different outputs. To quantify this, we use a variance
decomposition-based approach and calculate the commonly used Sobol
sensitivity indices \cite{sobol2001global}, considering first- and
total-order indices \cite{homma1996importance}. Methods for global
sensitivity analysis in most cases assume orthogonal (independent)
input factor distributions \cite{saltelli2008global}, and we have
implemented the Saltelli method \cite{saltelli2002making} with
guidance from \cite{halnes2009modelling} for this setting. In
\cite{eriksson2019uncertainty}, we also introduced the possibility of
calculating first-order sensitivity indices on the posterior
distribution obtained from an uncertainty quantification run
(Figure~\ref{fig:SA}), which in most cases is non-orthogonal, partly
following a scheme from \cite{saltelli2004sensitivity}. This approach
allows us, for example, to answer the following question: which
parameter, if known, would reduce the uncertainty in the predictions
the most? Answering this question can help guide further experiments.

\textbf{Parallelization} We adopt a \emph{simulation experiment} view
around using model solvers. We wrap all solver code around functions
that process simulation experiments (R lists) which can include data
alongside components like initial values, input parameters, output
time points. The \proglang{C} code component processes such
\emph{simulation experiments}, and the \proglang{R} wrapper uses
parallelization to process the experiments simultaneously. In addition
to this low level parallelization, we offer functions that process
several Markov chains within a parallel tempering scheme using
\texttt{pbdMPI}\footnote{\url{https://cran.r-project.org/package=pbdMPI}}
\cite{Chen2022pbdMPIpackage} .  This is a more coarse
parallelization, e.g.: the MPI analogue to the \texttt{mcmc} function
called \texttt{mcmc\_mpi}.  To use it, we write an R script with some
pbdMPI initialization boilerplate\footnote{necessary function calls
  like \texttt{init()} or \texttt{finalize()}}, and execute this file
in an MPI context, e.g.: \texttt{mpirun Rscript
  \emph{the-MPI-script.R}}. Several MPI workers (processes) are
started, each creating one Markov chain.  The MPI functions are
intended for use on a compute cluster. In the examples hosted on the
project's website, different MPI workers map to different parallel
tempering temperatures, but any choice of temperature is possible on
any MPI rank.

In our implementation of \emph{parallel tempering} the temperatures
themselves and the number of temperatures are not adaptive; we keep the
number of temperatures and the $\beta$ values themselves fixed. On
each Markov chain iteration, every chain offers to swap with every
other chain (in a sequence), using MPI broadcasts.

\subsection{Code examples and illustrations}
\label{sec:example}
At the documentation site
(\href{https://icpm-kth.github.io/uqsa/}{https://icpm-kth.github.io/uqsa/})
for \pkg{uqsa} we provide a range of different models and
corresponding experimental data to demonstrate the use of the
\pkg{uqsa} package. These examples are based on three biochemical
reaction networks, named AKAR4 \cite{church2021akap79}, AKAP79
\cite{buxbaum1989quantitative,church2021akap79}, and CaMKII
\cite{nair2014modeling,eriksson2019uncertainty} after some important
biochemical species in the respective networks. These networks have an
increasing order of complexity (size), where AKAR4 is the smallest
model, consisting of 4 species and 2 reactions, the AKAP79 model has
16 species and 16 reactions, and the CaMKII model 23 species and 36
reactions. The data is also a bit different, where AKAR4 and AKAP79
have time-series data and CaMKII dose response curves.

For each network, we provide an example on how to transfer this into a
deterministic (ODE) or a stochastic model and run a simulation. For
all networks and models, we also illustrate how to calibrate them with
UQ and SA using different algorithms.  On the package website, we
provide detailed \proglang{R} vignettes and articles with code
snippets that can be run on a laptop or desktop. The largest examples
should be run on a computing cluster.  The stochastic models are
simulated with our built-in Gillespie
algorithm\cite{GILLESPIE1976403} implementation. The deterministic
models in all of our examples use the built-in \proglang{C} back-end,
with solvers from the \texttt{odeiv2} module of the GNU Scientific
Library \cite{GSL}.

Below we provide code for AKAR4 but on the documentation site we also
provide examples with the larger models.

\textbf{Transferring a reaction model to a mathematical model and
  simulating.}

In a typical case, the user will already know where their TSV files
are stored, for the examples included with the package we call the
function \texttt{uqsa\_example} to locate the files.

In the following code snippet, we retrieve the location of the example
model (\texttt{f}: TSV file paths), load the model and data
(\texttt{m}: list of data frames), interpret the reaction network
model as an ODE (\texttt{o}: list), compile auto-generated
\proglang{C} code for this ODE model (shared library), read simulation
instructions for the different experiments (\texttt{ex}):
\begin{CodeChunk}
\begin{CodeInput}
library(uqsa)
library(parallel)
opt <- options(mc.cores=parallel::detectCores())
f <- uqsa_example("AKAR4")
m <- model_from_tsv(f)
o <- write_and_compile(as_ode(m))
ex <- experiments(m,o)
\end{CodeInput}
\end{CodeChunk}
The ODE is written as C-code to a temporary location and compiled
there (\texttt{write\_and\_compile()}), the path of the compiled
shared library\footnote{e.g.:
  \texttt{/tmp/RtmpkIYXUe/AKAR4\_bb0b51606d02.so}} is saved in the
variable \texttt{o} for convenience. The last line
\texttt{experiments(m,o)} extracts the experimental data from the
tsv-files, while taking the ODE structure and conservation laws into
account. This returns a list of simulation experiments, with initial
values for the ODE, input parameters (per experiment), and the
measured data (as objects of class \texttt{errors} and as plain
data-frames):
\begin{CodeChunk}
\begin{CodeInput}
print(names(ex))
\end{CodeInput}
\begin{CodeOutput}
[1] "400nM" "100nM" "25nM"
\end{CodeOutput}
\begin{CodeInput}
print(class(ex))
\end{CodeInput}
\begin{CodeOutput}
[1] "experiments"
\end{CodeOutput}
\begin{CodeInput}
print(ex)
\end{CodeInput}
\begin{CodeOutput}
number of simulation experiments: 3
                                     400nM
------------------------------------------
            measurements: 2 columns (data.frame)
                    data: 1, 225 (dim)
                   input: 2 (length)
             initialTime: -15
            initialState: 2 (length)
             outputTimes: 225 (length)
                  events: NULL (class), NULL (type)

                                     100nM
------------------------------------------
            measurements: 2 columns (data.frame)
                    data: 1, 225 (dim)
                   input: 2 (length)
             initialTime: -15
            initialState: 2 (length)
             outputTimes: 225 (length)
                  events: NULL (class), NULL (type)

                                      25nM
------------------------------------------
            measurements: 2 columns (data.frame)
                    data: 1, 225 (dim)
                   input: 2 (length)
             initialTime: -15
            initialState: 2 (length)
             outputTimes: 225 (length)
                  events: NULL (class), NULL (type)

experiments:  400nM, 100nM, 25nM
\end{CodeOutput}
\begin{CodeInput}
print(ex[[1]][['input']])
\end{CodeInput}
\begin{CodeOutput}
AKAR4_C_ConservedConst   AKAR4_ConservedConst
                   0.4                   -0.2
\end{CodeOutput}
\end{CodeChunk}

\textbf{Parameter estimation and uncertainty quantification} The below
code snippet creates a closure \texttt{s}, to simulate the
mathematical model. The closure implicitly stores the experiments
(\texttt{ex}) to be simulated (including the measured data), so that
this doesn't have to be stated on every simulation:
\begin{CodeChunk}
\begin{CodeInput}
s <- simulator.c(ex,o,parMap=log10ParMap,omit=2)
p <- log10(values(m[['Parameter']]))
y <- s(p)
\end{CodeInput}
\begin{CodeInput}
class(y)
\end{CodeInput}
\begin{CodeOutput}
[1] "simulation"
\end{CodeOutput}
\begin{CodeInput}
print(y)
\end{CodeInput}
\begin{CodeOutput}
number of simulation experiments: 3
                                     400nM
------------------------------------------
              cpuSeconds: 0.000445
                numSteps: 269
                  status: 0
                   state: 2, 225, 1 (dim)
                    func: 1, 225, 1 (dim)
           logLikelihood: -864.384
       gradLogLikelihood: NULL (class), NULL (type)
       FisherInformation: NULL (class), NULL (type)

                                     100nM
------------------------------------------
              cpuSeconds: 0.000447
                numSteps: 254
                  status: 0
                   state: 2, 225, 1 (dim)
                    func: 1, 225, 1 (dim)
           logLikelihood: -840.397
       gradLogLikelihood: NULL (class), NULL (type)
       FisherInformation: NULL (class), NULL (type)

                                      25nM
------------------------------------------
              cpuSeconds: 0.000287
                numSteps: 236
                  status: 0
                   state: 2, 225, 1 (dim)
                    func: 1, 225, 1 (dim)
           logLikelihood: -786.464
       gradLogLikelihood: NULL (class), NULL (type)
       FisherInformation: NULL (class), NULL (type)
\end{CodeOutput}
\end{CodeChunk}
Next, we create a sampling closure, which can create samples from the
posterior distribution, using the Metropolis algorithm (in this case):
\begin{CodeChunk}
\begin{CodeInput}
sd <- rep(2,length(p))
dprior <- dNormalPrior(p,sd)
rprior <- rNormalPrior(p,sd)
mh <- mcmc(
	metropolis_update(
		s,
		dprior=dprior,
		Sigma=diag(sd^2)
	)
)
\end{CodeInput}
\end{CodeChunk}
Here, we obtain a sample of size \texttt{N}, with step size \texttt{h}:
\begin{CodeChunk}
\begin{CodeInput}
set.seed(1337)
x <- mcmc_init(beta=1.0,p,s,ll,dprior)
h <- tune_step_size(mh,x,h=0.1)
N <- 1e4
t_initial <- Sys.time()
mediumSizedSample <- mh(x,N,h)
t_final <- Sys.time()
cat("wall clock time: ",difftime(t_final, t_initial,units="min")," minutes\n")
colnames(mediumSizedSample) <- names(p)
\end{CodeInput}
\begin{CodeOutput}
wall clock time:  2.618143  minutes
\end{CodeOutput}
\end{CodeChunk}
To plot this sample, we want to omit some points in the sample, to reduce the plot complexity:
\begin{CodeChunk}
\begin{CodeInput}
S_ <- small(mediumSizedSample)
L_ <- attr(S_,"logLikelihood")
Y_ <- s(t(S_))
if (requireNamespace("hexbin",quietly=TRUE)){
	hexbin::hexplom(S_,xbins=10)
} else {
	showPosterior(S_,rprior(NROW(S_)))
}
\end{CodeInput}
\begin{CodeOutput}
Auto-correltation length:  4
Loading required package: hexbin
Loading required namespace: ks
\end{CodeOutput}
\end{CodeChunk}

\textbf{Model the posterior with \pkg{VineCopula}} From the posterior
sample above, we can create an inferred density function, using the
\texttt{VineCopula} package:
\begin{CodeChunk}
\begin{CodeInput}
C <- fitCopula(S_)
rposterior <- rCopulaPrior(C)
dposterior <- dCopulaPrior(C)
\end{CodeInput}
\end{CodeChunk}
These two closures can be used in exactly the same way as
\texttt{rprior} and \texttt{dprior}. Once new data is measured, the
Bayesian analysis can be iterated, with \texttt{\{r,d\}posterior}
taking the slots that \texttt{\{r,d\}prior} have occupied\footnote{the
  current posterior distribution is the prior distribution of the next
  iteration}:
\begin{CodeChunk}
\begin{CodeInput}
rprior(4)
\end{CodeInput}
\begin{CodeOutput}
           [,1]      [,2]      [,3]
[1,] -2.8368493 -2.512971 1.9199596
[2,] -3.4880754 -3.759134 0.8668123
[3,] -0.8662677  1.935870 0.6495512
[4,] -2.6190208 -1.135678 2.667794
\end{CodeOutput}
\begin{CodeInput}
rposterior(4)
\end{CodeInput}
\begin{CodeOutput}
          [,1]         [,2]      [,3]
[1,] -1.717124 -0.005826695 4.2204216
[2,] -1.737549 -0.352981031 1.7643233
[3,] -1.733924 -1.585463243 1.3407122
[4,] -1.583772 -1.254826988 0.5231776
\end{CodeOutput}
\end{CodeChunk}
The names the functions that produces these closures
\texttt{\{r,d\}CopulaPrior()} (\emph{Prior}) imply that the foremost
reason to use them is to construct a prior distribution from earlier
Bayesian analysis. But it is of course not the only purpose.

The function \texttt{dposterior} is an approximation of the posterior
density function and \texttt{rposterior} produces samples from this
approximation, similar to the Markov chain simulation of this
distribution, except that it is limited to the shape that was
discovered by MCMC (or other methods of that kind). Let us assume that
we explored the entire posterior density, and now we compare
\texttt{rposterior} to the sample it is based on. To make two
dimensional plots we redo the copula fit on the first two sample
variables to estimate the marginal distribution:
\begin{CodeChunk}
\begin{CodeInput}
C_2D <- fitCopula(S_[,c(1,2)])
rposterior <- rCopulaPrior(C_2D)
dposterior <- dCopulaPrior(C_2D)
\end{CodeInput}
\end{CodeChunk}
Next we produce a two dimensional sample (\texttt{Z\_}) from the
marginal distribution (\texttt{rposterior}), and calculate density values on a grid of
points (\texttt{dposterior}):
\begin{CodeChunk}
\begin{CodeInput}
Z_ <- rposterior(NROW(S_))
x <- quantile(S_[,1],probs=seq(0,0.95,length.out=100))
y <- quantile(S_[,2],probs=seq(0,0.95,length.out=100))
z <- outer(x,y,function(x,y) dposterior(cbind(x,y)))
b <- 2^(seq(0,ceiling(log2(max(z)))))
dev.new()
oldpar <- par(mfrow=c(3,1),bty='n')
LIM <- list(
	x=quantile(S_[,1],probs=c(0,0.95)),
	y=quantile(S_[,2],probs=c(0,0.95))
)
plot(
	S_[,1],S_[,2],
	type='p',
	xlab=colnames(S_)[1], ylab=colnames(S_)[2],
	xlim=LIM[['x']], ylim=LIM[['y']]
)
plot(
	Z_[,1],Z_[,2],
	type='p',
	xlab=colnames(S_)[1],ylab=colnames(S_)[2],
	xlim=LIM[['x']],ylim=LIM[['y']]
)
contour(
	x,y,z,
	xlab=colnames(S_)[1],ylab=colnames(S_)[2],
	drawlabels=TRUE,
	xlim=LIM[['x']],ylim=LIM[['y']],
	levels=b
)
par(oldpar)
\end{CodeInput}
\end{CodeChunk}

\textbf{Global sensitivity analysis} And finally, we estimate the
sensitivity of the output function \texttt{AKAR4p} with respect to
each of the three parameters, at each time-point:
\begin{CodeChunk}
\begin{CodeInput}
GSA <- gsa_binning(S_,t(Y_[[1]][['func']]["AKAR4pOUT",,]))
tm <- ex[[1]][['outputTimes']]
sensitivity.graph(
	tm,GSA,
	xlab="time",ylab="sensitivity",
	main="simulation experiment 1"
)
\end{CodeInput}
\end{CodeChunk}
Here we used experiment 1 as an example, Figure~\ref{fig:SA} we show another example with an unobserved state and and experiment.
\begin{figure}
  \centering\hspace*{-1cm}
  \includegraphics[width=1\linewidth]{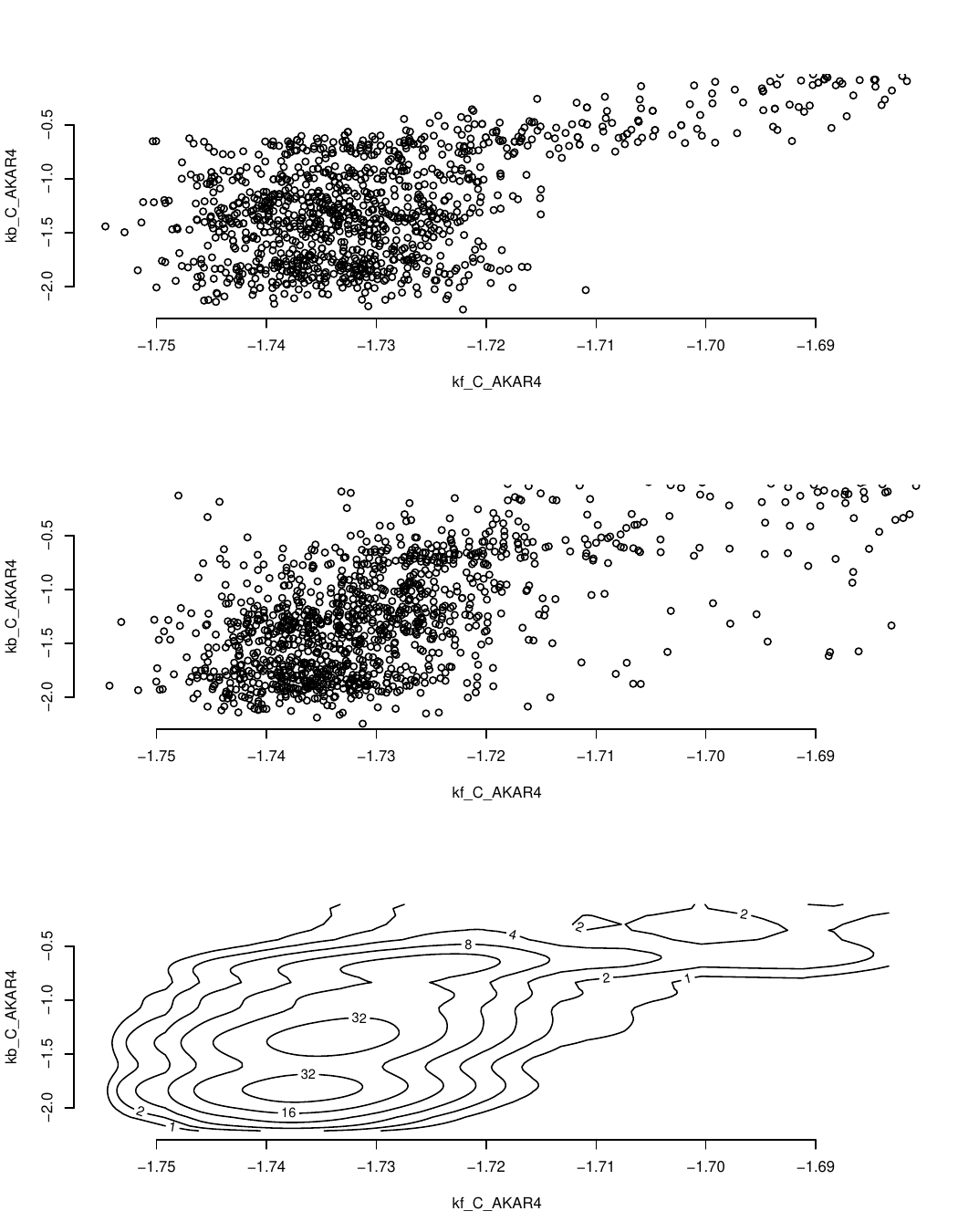}
  \caption[AKAR4 2d posterior]{2D projection of the AKAR4 posterior
    (roughly inferred from short MCMC). \emph{top}: The sample from
    MCMC, \emph{mid}: a sample from the VineCopula approximation,
    \emph{bottom}: the contour lines of the density function (via
    VineCopula package). Two variables were allowed to vary, while the
    rest (one in this case) were fixed to the maximum-likelihood
    value.}
  \label{fig:AKAR4-POST}
\end{figure}
And finally we restore the original options:
\begin{CodeChunk}
\begin{CodeInput}
options(opt)
\end{CodeInput}
\end{CodeChunk}

\begin{figure}
  \centering
  \includegraphics[width=\textwidth]{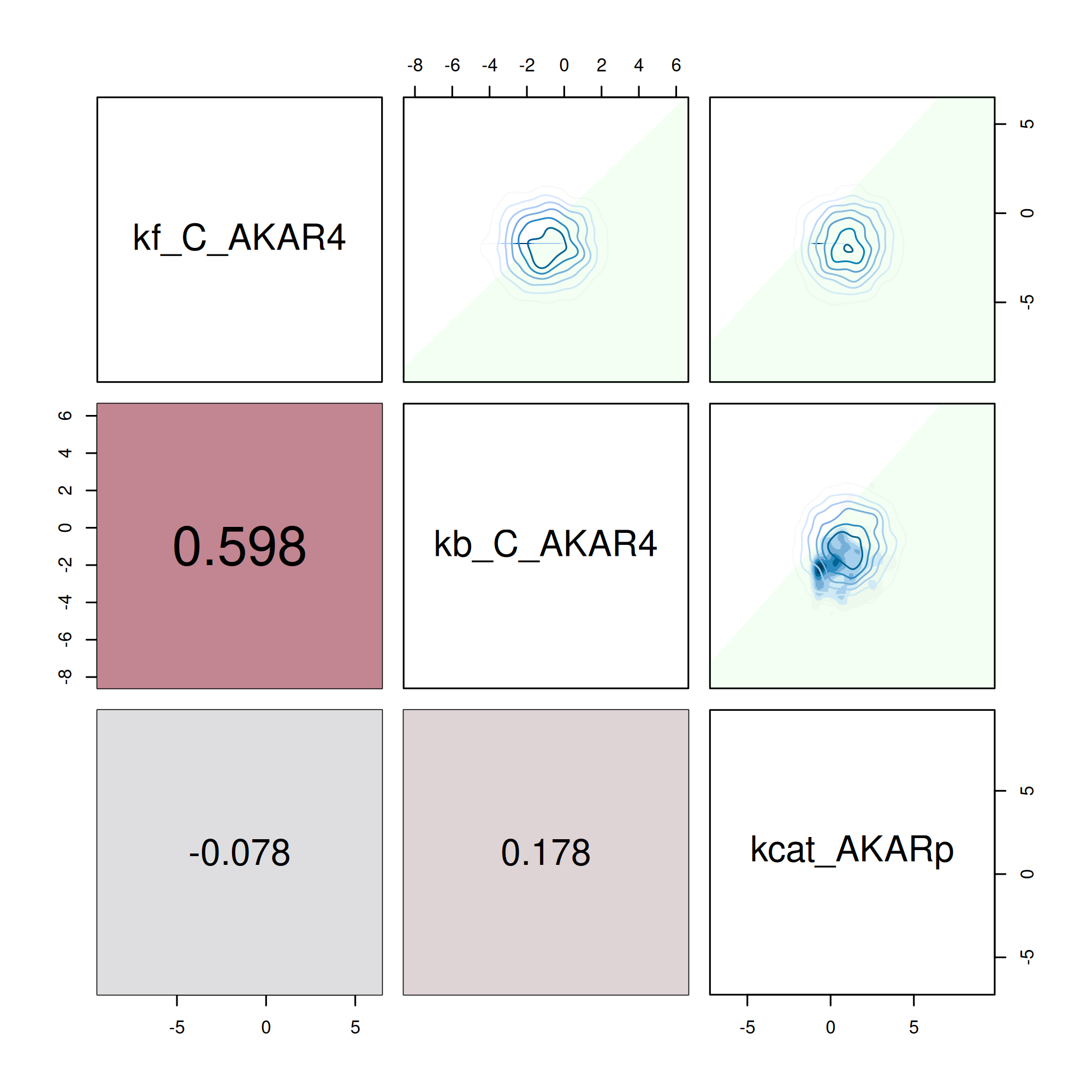}
  \caption{A pairwise plot of the AKAR4 sample. shaded areas represent
    the posterior, contour-lines represent the prior
    distribution.\label{fig:AKAR4-posterior}}
\end{figure}

\section{Discussion}

The \pkg{uqsa} tool addresses the need for improved validation of
biochemical network models arising, for example, in neuroscience and
systems biology. With the \pkg{uqsa} package, the uncertainty in
parameter estimates can be quantified at the same time as parameter
estimation is performed. This uncertainty can also be propagated to
model predictions, and a global sensitivity analysis can be carried
out to guide the design of further experiments. Tools for uncertainty
quantification exist in other languages (see, e.g.,
pyABC~\cite{schaelte2022pyabc}), our \proglang{R} toolbox introduces
several distinctive features: seamless integration with other R
packages (mclust, VineCopula) and offers not only uncertainty
quantification but also global sensitivity analysis functionality. It
is tailored to biochemical reaction network models, but allows any ODE
model as long as it is compatible with GSL solvers. We see it as a
distinctive advantage that we allow the user to choose between two
different modeling strategies for the same reaction network model: the
user can freely switch between deterministic ODEs and fully stochastic
CMEs. \pkg{uqsa} is fast and easy-to-use.

A variety of methodologies is available for parameter estimation and
uncertainty quantification. We chose the Bayesian framework to address
the UQ problem, quantifying the uncertainty through the posterior
distribution of the parameters. Within this framework, we rely on MCMC
methods, which are particularly well suited to this class of problems,
and relatively straightforward to implement. However, MCMC sampling,
whether in the likelihood-free (ABC) or likelihood-based setting,
comes with substantial computational costs. MCMC methods are typically
based on a mechanism of proposing samples and accepting or rejecting
them. Each proposed sample requires a model simulation, even if the
proposal is ultimately rejected. This process of proposing,
simulating, and accepting/rejecting samples must be repeated many
thousands of times. It is therefore essential to employ fast and
efficient ODE and stochastic solvers in the simulation step, in order
to prevent the sampling method from becoming prohibitively expensive.

Sampling explores the volume of the parameter space within the limits
of acceptable fits, and therefore scales exponentially with the number
of parameters\footnote{This is more demanding than optimization - a
  large optimization problem differs from a large sampling problem by
  orders of magnitude.}. In addition, MCMC samples have intrinsic
(integrated) auto-correlation, so that the \emph{effective sample
  size} is reduced when auto-correlation is high,
see~\cite{ballnus2017}. Poor MCMC configurations (transition kernel
choice, etc.) can therefore lead to samples with large
auto-correlations. Furthermore, MCMC methods typically require a good
initial point to begin sampling and a convergence phase, during which
the target distribution is approached. Several approaches have been
developed to specifically alleviate these difficulties in a general
way: parallel tempering~\cite{Earl2005}, Hamiltonian Monte
Carlo~\cite{Betancourt2017}, Fisher information as metric tensors for
movement~\cite{Girolami2011}, and adaptive tuning of acceptance
rates~\cite{Bedard2008}. However, for some model classes, including
ordinary differential equation models, these approaches can in
themselves increase the computational costs substantially. Thus, the
urgent need for numerically efficient implementations remains,
regardless of the chosen
algorithm~\cite{eriksson2021sensitivity}. Other approaches, such as
the profile likelihood method~\cite{Kreutz2013}, address the scaling
problem directly by exploring the shape of the likelihood (in profile)
rather than sampling from a target distribution.

For GSA, a range of methods is available \cite{zi2011sensitivity}. For
example, Tenn{\o}e et. al. (2018) \cite{tennoe2018uncertainpy}
developed an extensive Python package focusing on GSA for neuroscience
models. To our knowledge, however, \pkg{uqsa} is the only package that
combines Bayesian UQ and SA within a single framework.

Uncertainty quantification is not only necessary for validating (or
falsifying) a given model structure, but is also important for the
reproducibility of the model calibration process
\cite{eriksson2022combining}. When optimization is used to identify a
single best parameter value, different methods may produce different
but equally good solutions since there can be several local minima. As
a consequence, the outcome may not be fully reproducible. In contrast,
a Bayesian UQ approach should, at least in principle, sample the full
volume of viable parameter sets, and can therefore be compared more
meaningfully across methods.

It is important to note, however, that \pkg{uqsa} quantifies only
parameter uncertainty. Uncertainty in the model structure itself is
not yet addressed. We aim to incorporate this aspect in future
versions of the \pkg{uqsa} tool.

\section*{Acknowledgments}
We thank Karl Johan Westrin and Henric Zazzi at PDC Centre for High
Performance Computing, KTH Royal Institute of Technology, for help
with pbdMPI. The computations were enabled by resources provided by
the National Academic Infrastructure for Supercomputing in Sweden
(NAISS) and the Swedish National Infrastructure for Computing (SNIC)
at PDC Centre for High Performance Computing, partially funded by the
Swedish Research Council through grant agreements no. 2022-06725 and
no. 2018-05973 and by LUNARC, The Centre for Scientific and Technical
Computing at Lund University.

\section*{Funding}
The research was supported by Swedish e-Science Research Centre (SeRC)
and Science for Life Laboratory. AK and OE were supported by the
Swedish Foundation for Strategic Research (SSF), grant
no. ITM24-0470. The research of AK, OE and JHK was supported by
EU/Horizon 2020 no. 945539 (HBP SGA3), the EU/Horizon 2020
no. 101147319 (EBRAINS 2.0 Project), European Union's Research and
Innovation Program Horizon Europe no 101137289 (the Virtual Brain Twin
Project) and the Swedish Research Council (VR-M-2020-01652).

\bibliographystyle{alpha}
\bibliography{main}


\newpage


\begin{tabular}{l}
  Federica Milinanni\\
  Institute for Computational and Experimental Research in Mathematics (ICERM)\\
  Brown University,\\
  Providence, RI 02903\\
  USA\\
  E-mail: \email{federica\_milinanni@brown.edu}\\
  URL: \url{https://sites.google.com/view/federicamilinanni}
  \\
  \\
  Olivia Eriksson\\
  Science for Life Laboratory\\
  School of Electrical Engineering and Computer Science\\
  KTH Royal Institute of Technology\\
  SE-100 44 Stockholm\\
  Sweden\\
  E-mail: \email{olivia@kth.se}\\
  URL: \url{https://www.kth.se/profile/olivia}
  \\
  \\
\end{tabular}

\end{document}